\documentclass[sigconf]{acmart}

\AtBeginDocument{%
  }

\copyrightyear{2024}
\acmYear{2024}
\setcopyright{rightsretained}
\acmConference[RecSys '24]{18th ACM Conference on Recommender Systems}{October 14--18, 2024}{Bari, Italy}
\acmBooktitle{18th ACM Conference on Recommender Systems (RecSys '24), October 14--18, 2024, Bari, Italy}\acmDOI{10.1145/3640457.3691701}
\acmISBN{979-8-4007-0505-2/24/10}

\begin{document}

\title{\textit{RePlay}: a Recommendation Framework for Experimentation and Production Use}

\author{Alexey Vasilev}
\email{alexxl.vasilev@yandex.ru}
\orcid{0009-0007-1415-2004}
\affiliation{
  \institution{Sber AI Lab}
  \city{Moscow}
  \country{Russian Federation}
}

\author{Anna Volodkevich}
\email{volodkanna@yandex.ru}
\orcid{0009-0002-7958-0097}
\affiliation{
  \institution{Sber AI Lab}
  \city{Moscow}
  \country{Russian Federation}
}

\author{Denis Kulandin}
\email{kulandin08@gmail.com}
\orcid{0009-0001-6338-4190}
\affiliation{
  \institution{Sber}
  \city{Nizhnii Novgorod}
  \country{Russian Federation}
}

\author{Tatiana Bysheva}
\email{bysheva.tatyana@gmail.com}
\orcid{0009-0002-3758-6482}
\affiliation{
  \institution{Sber}
  \city{Nizhnii Novgorod}
  \country{Russian Federation}
}

\author{Anton Klenitskiy}
\email{antklen@gmail.com}
\orcid{0009-0005-8961-6921}
\affiliation{
  \institution{Sber AI Lab}
  \city{Moscow}
  \country{Russian Federation}
}

\renewcommand{\shortauthors}{Alexey Vasilev et al.}

\begin{abstract}
  Using a single tool to build and compare recommender systems significantly reduces the time to market for new models. In addition, the comparison results when using such tools look more consistent. This is why many different tools and libraries for researchers in the field of recommendations have recently appeared. Unfortunately, most of these frameworks are aimed primarily at researchers and require modification for use in production due to the inability to work on large datasets or an inappropriate architecture. In this demo, we present our open-source toolkit RePlay - a framework containing an end-to-end pipeline for building recommender systems, which is ready for production use. RePlay also allows you to use a suitable stack for the pipeline on each stage: Pandas, Polars, or Spark. This allows the library to scale computations and deploy to a cluster. Thus, RePlay allows data scientists to easily move from research mode to production mode using the same interfaces.
\end{abstract}



\maketitle

\section{Introduction}

The field of recommender systems has been actively developing in recent decades, with new algorithms and approaches constantly emerging and improving. Researchers also increasingly pay attention to the issue of reproducibility and verification of certain methods and their correct evaluation \cite{bench,isseq}.

Due to the increase and diversity of various recommender algorithms, researchers and engineers from the industry have difficulties with reproducing and comparing recommender algorithms since, in scientific articles, the code provided by the authors may not work or be absent altogether. Other stages of the recommender pipeline, such as calculating metrics, splitting, and preprocessing, may also differ. That is why several frameworks for recommendations have appeared in the last few years, such as \cite{recpack,eliot,recbole,msrec}. They make the process of comparing recommender algorithms more universal and allow to avoid simple errors when comparing, such as different implementations of metrics in the compared algorithms \cite{tamm21}.

\begin{figure*}[ht!]

\centering
  \includegraphics[width=.99\textwidth]{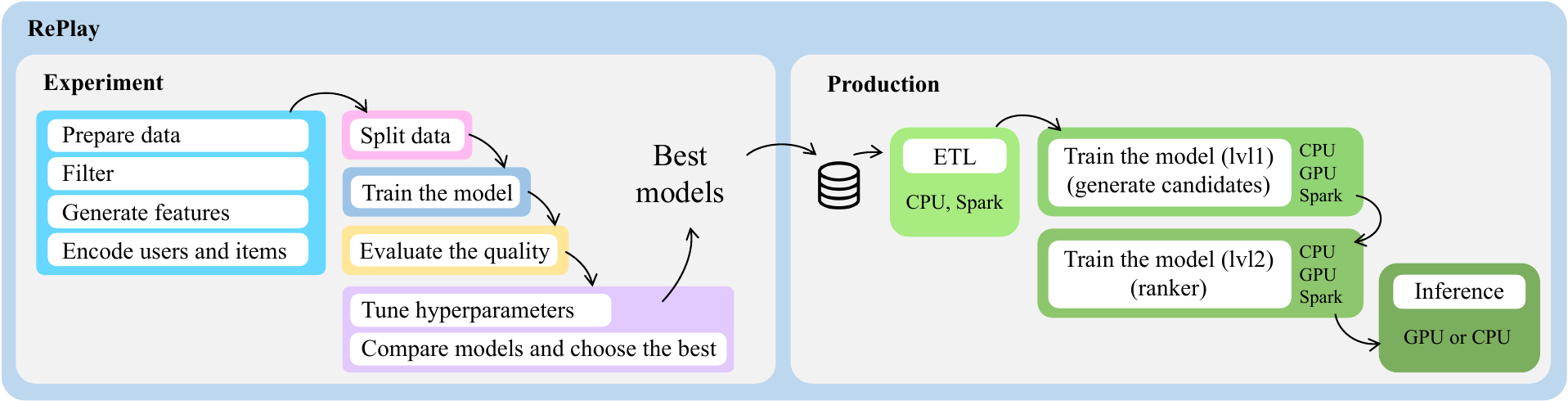}
  \caption{RePlay pipeline. Using the RePlay framework, it is possible to implement both experimentation and production pipelines.}
  \label{fig:replay}
  \vspace{-0.4cm}
\end{figure*}

Such libraries usually contain a fairly large set of algorithms and metrics for comparison. However, most of these frameworks are intended specifically for researchers since they can work quite slowly, or the code will require revision for implementation in production. We present \textit{RePlay} - a framework that contains all stages of the recommendation pipeline but is more focused on use in production. Our library is an experimentation and production toolkit for top-N recommendation. \textit{RePlay} has rich test coverage and detailed documentation.

\textit{RePlay} supports three types of dataframes: Spark, Polars, and Pandas, as well as different types of hardware architecture: CPU, GPU, and cluster, so you can choose a convenient configuration on each stage of the pipeline depending on the model and your hardware. In addition, many basic models are written in Spark or are wrappers of Spark implementations, which makes it easy to scale computations and deploy to a cluster. Of the frameworks described above, only Recommenders \cite{msrec} support both Spark and Pandas. 

Figure \ref{fig:replay} shows the experimentation and production pipelines with RePlay.

The main features of \textit{RePlay} are the following: 
\begin{enumerate}
\item Production ready code, which can be embedded in your recommendation platform
\item Possible to implement both experimentation and production pipelines
\item Support for various types of dataframes: Spark, Polars, Pandas
\item Support for various types of hardware architectures: CPU, GPU, Cluster
\end{enumerate}

In this demo, we will give you an overview of the various \textit{RePlay} modules and how to use them for experimentation and production. \textit{RePlay} source code is available on GitHub\footnote{https://github.com/sb-ai-lab/RePlay}.

\section{RePlay}

The main components of the library are briefly described below.

\subsubsection*{Preprocessing}

The preprocessing module contains different filters which can be applied to a dataframe with interactions. For example, the \textit{MinCountFilter} filters out users or items presented less than a given number of times. The \textit{LowRatingFilter} filters out interactions with low ratings or relevance values. The \textit{TimePeriodFilter} allows you to select the required period of time. Several filters select a given number of days or interactions globally or by users.

\subsubsection*{Splitters}

The splitters module provides various strategies for splitting a dataframe with interactions into train and test parts. All splitters include options for dropping cold users and items not presented in the train data from the test data. The simple \textit{RandomSplitter} assigns records into train and test at random. The \textit{ColdUserRandomSplitter} makes split by users, so the test set consists of all actions of randomly chosen users. The \textit{TimeSplitter} is responsible for classic split by time, while the \textit{LastNSplitter} takes the last $N$ interactions for each user into a test set. The \textit{NewUsersSplitter} allows to assign only new cold users to the test set.

\subsubsection*{Data}

RePlay utilizes a special \textit{Dataset} class to store all preprocessed data needed for training models. It acts as a container for the interactions dataframe as well as the item and user features. An instance of the \textit{Dataset} class requires dataframes with a fixed schema described by the \textit{FeatureSchema} class. The \textit{FeatureSchema} contains information about all dataset features, including column name, feature type, the source dataframe this feature comes from, the cardinality of the feature, and feature hint (whether this feature corresponds to item id, user id, timestamp, or rating).

For many algorithms, it is necessary to encode user and item ids as consecutive integers without any gaps. The \textit{DatasetLabelEncoder} performs such encoding and stores corresponding mappings, allowing to convert ids back after model prediction. Additionally, it handles conversions for all other categorical features in the dataset. Due to the presence of the \textit{FeatureSchema}, all encoding is done automatically, eliminating the need for any configuration.

\subsubsection*{Models}

RePlay provides a wide range of recommendation algorithms. There are several popularity-based baseline models - \textit{PopRec}, \textit{QueryPopRec}, and \textit{Wilson} recommenders. Classic collaborative filtering algorithms are represented by item-based nearest neighbors (\textit{ItemKNN}), matrix factorization with Alternating Least Squares (\textit{ALSWrap}), and \textit{SLIM} \cite{ning2011slim}. In addition, there are recommenders based on Word2Vec (\textit{Word2VecRec}) and clustering-based (\textit{ClusterRec}). The item-to-item model \textit{AssociationRulesItemRec} is based on association rules. For factorization machines, there is a wrapper for the lightfm library \cite{DBLP:conf/recsys/Kula15} - \textit{LightFMWrap}. Another wrapper is \textit{ImplicitWrap} for the popular implicit\footnote{\url{https://github.com/benfred/implicit}} library. There are a couple of deep learning-based collaborative algorithms - \textit{NeuroMF} \cite{he2017neural} for Neural Matrix Factorization and \textit{MultVAE} \cite{liang2018variational} for autoencoders. The reinforcement learning family of models is represented by classic bandit approaches (\textit{UCB} \cite{auer2002finite} and \textit{KLUCB} \cite{garivier2011kl} recommenders, \textit{ThompsonSampling} \cite{chapelle2011empirical}) and more sophisticated models - \textit{DDPG} \cite{liu2018deep}, Decision Transformer (\textit{DT4Rec} \cite{zhao2023user}) and Conservative Q-Learning \cite{xiao2021general} (\textit{CQL} recommender). Finally, RePlay includes two state-of-the-art models for sequential recommendations - SASRec \cite{kang2018self} and BERT4Rec \cite{sun2019bert4rec}. Following current best practices \cite{klenitskiy23, petrov2023gsasrec, wilm2023scaling}, it is possible to train these models with different types of loss functions (cross-entropy or binary cross-entropy) and different negative sampling strategies.

Deep learning-based models are implemented with PyTorch. Other algorithms are implemented with PySpark and are ready for large-scale distributed computing. Most models follow a common scikit-learn style interface with \textit{fit} and \textit{predict} methods. Sequential deep learning models (SASRec and BERT4Rec) use pytorch-lightning\footnote{\url{https://github.com/Lightning-AI/pytorch-lightning}} framework for training and inference and may be applied for both online and offline inference.

\subsubsection*{Hyperparameter tuning}

Model classes have a method \textit{optimize} to perform hyperparameter tuning with the popular and flexible Optuna library \cite{optuna}.

\subsubsection*{Metrics}

The metrics module is responsible for model evaluation. It contains the most widely used recommendation metrics. Among them are Precision, Recall, Mean Average Precision (MAP), Mean Reciprocal Rank (MRR), Normalized Discounted Cumulative Gain (NDCG), Hit Rate, and ROC AUC. Several well-known beyond-accuracy metrics (Categorical Diversity, Coverage, Novelty, Surprisal, and Unexpectedness) are implemented as well. The \textit{OfflineMetrics} class provides a more efficient way to calculate multiple metrics for the same input data simultaneously. The \textit{Experiment} class is designed for performance comparison of different models or hyperparameters. It collects metrics from different runs. In addition, it is possible to add a custom metric, which should inherit from the base \textit{Metric} class, and implement the \textit{\_get\_metric\_value\_by\_user} method with calculation for one user.

\section{Setup}

RePlay library is available for installation from the site \href{https://pypi.org/project/replay-rec/}{\texttt{pypi.org}} under the name \textit{replay-rec}. The core package without PySpark and PyTorch dependencies will be installed by default. For convenience, the PySpark and PyTorch functionality can be installed separately as additional extras. For more details, see the \href{https://github.com/sb-ai-lab/RePlay/blob/main/README.md\#installation}{\textit{Installation}} section in the Repository or the \href{https://sb-ai-lab.github.io/RePlay/index.html}{\textit{Library Documentation}}. 

Some RePlay functionality is available in the \textit{experimental} package. The core package requires a limited number of dependencies to be easily implemented into production. The models from the \textit{experimental} package could differ from the core package models in APIs. The experimental package does not have strict test coverage requirements as the core one to allow fast experiments. The experimental package, indicated with the \textit{rc0} suffix, can also be installed from PyPI.

To raise issues or ask questions about the use of RePlay, check out the source code and \href{https://github.com/sb-ai-lab/RePlay/blob/main/CONTRIBUTING.md}{\textit{Contribution Guidelines}} on GitHub.

\section{Demo}

In the demo, we demonstrate the main stages of the RePlay pipeline using the MovieLens 1M dataset \cite{harper2015movielens}. First, we apply a leave-one-out split with the \textit{LastNSplitter}. Next, we load the data into the \textit{Dataset} class, standard for all RePlay models, and transform it with the \textit{SequenceTokenizer} to store the data as sequences. Further, we train the SASRec model and measure the MAP, NDCG, and Recall on the validation data. After that, we get the recommendations in different data types: PySpark, Pandas, Polars dataframes, and PyTorch Tensors. Finally, we calculate various metrics for the different K on the test data with the \textit{OfflineMetrics} class. 

\section{Conclusion and future work}

RePlay provides all the essential pipeline steps for recommender system researchers and developers. It allows you to conduct experiments with Pandas or Polars and facilitates a smooth transition to Spark for large-scale computations in production.
RePlay is actively maintained and developed. In the near future, we plan to expand the number of neural network algorithms, as well as basic algorithms implemented on Polars.

\section{Acknowledgements}
We wish to express our sincere gratitude to the people, who have made significant contributions to the development of RePlay: Boris Shminke, Yan-Martin Tamm, Alexey Grishanov, Nikolay Butakov, Egor Bodrov, Eduard Malov, Maxim Savchenko, and Alexander Tuzhilin. Additionally, we extend our heartfelt thanks to Alexandra Laricheva for her valuable assistance with the auxiliary materials for our demo.

\bibliographystyle{ACM-Reference-Format}
\bibliography{main}
\end{document}